\documentclass[aps,prl,reprint,floatfix]{revtex4-1}
\usepackage{blindtext}
\usepackage{lipsum}  
\usepackage{amsmath}
\usepackage{amsfonts}
\usepackage{amssymb}
\usepackage{comment}
\usepackage{graphicx}
\usepackage{dcolumn}
\usepackage{bm}
\usepackage{physics}
\usepackage[normalem]{ulem}
\usepackage{xcolor}

\begin{document}
\title{Viscoelastic confinement induces periodic flow reversals in active nematics}
\author{Francesco Mori$^1$, Saraswat Bhattacharyya$^1$, Julia M. Yeomans$^1$, Sumesh P. Thampi$^2$}
 \affiliation{$^1$ Rudolf Peierls Centre for Theoretical Physics, University of Oxford, Oxford OX1 3PU, United Kingdom\\ $^2$  Department of Chemical Engineering, Indian Institute of Technology Madras, Chennai-36, India}

\begin{abstract}

We use linear stability analysis and hybrid lattice Boltzmann simulations to study the dynamical behaviour of an active nematic confined in a channel made of viscoelastic material. We find that the quiescent, ordered active nematic is unstable above a critical activity. The transition is to a steady flow state for high elasticity of the channel surroundings. However, below a threshold elastic modulus, the system produces spontaneous oscillations with periodic flow reversals. We provide a
phase diagram that highlights the region where time-periodic oscillations are observed and explain how they are produced by the interplay of activity and viscoelasticity. Our results suggest new experiments to study the role of viscoelastic confinement in the spatio-temporal organization and control of active matter.
\end{abstract}

\maketitle

Living systems across scales exhibit collective motion, and thus spatio-temporal patterns, vividly manifested as, for instance, motility-induced phase separation \cite{cates2015motility}, spontaneous flow transitions \cite{voituriez2005spontaneous,edwards2009spontaneous,giomi2012banding,duclos2018spontaneous}, and turbulence at low Reynolds number \cite{marchetti2013hydrodynamics,thampi2016active,alert2022active}. Not only biochemical and genetic cues but mechanical interactions of the system with its surroundings are important in dictating such emergent dynamics. Adding to this complexity, biological environments are often endowed with viscoelastic properties, for example, biofilms where bacterial cells colonize in a polymeric matrix \cite{conrad2018confined}, migration of cells through extracellular matrix \cite{friedl2009collective,chaudhuri2020effects,clark2022self,elosegui2023matrix}, notably the phenomenon of durotaxis \cite{sunyer2020durotaxis}, and change in the swimming behaviour of microorganisms due to the presence of polymers in biofluids \cite{patteson2015running,zottl2019enhanced}. In a different context, traction force microscopy has become an indispensable tool to probe force fields in cellular structures. The technique assumes a one-way mechanical interaction of cells with an elastic substrate \cite{style2014traction,colin2018future}. Therefore, clarifying the interplay of the viscoelasticity of a confining medium and activity of the living system is crucial from understanding measurements in mechanobiology to biological events such as wound healing \cite{brugues2014forces}, morphogenesis \cite{chiou2018we}, and cancer invasion \cite{weigelin2012intravital}. Besides, identifying universal pathways of pattern formation is a central goal of active matter research.

It is well known that active nematics, a versatile model fluid for active matter, confined in a rigid channel displays a transition---mathematically analogous to the Fredericks transition in passive liquid crystals \cite{de1993physics}---from quiescence to a flow state when the activity is increased beyond a threshold value \cite{voituriez2005spontaneous,deforet2014emergence,wioland2016directed,chandrakar2020confinement,singh2023three}. Further increase in activity induces a cascade of dynamical transitions resulting in oscillatory flows \cite{giomi2012banding,hardouin2019reconfigurable}, dancing topological defects \cite{shendruk2017dancing,hardouin2019reconfigurable}, and active turbulence \cite{thampi2022channel,opathalage2019self,chandragiri2019active,samui2021flow,joshi2023disks}. Thus, channel-confined active nematics have become a paradigm for understanding the dynamical behaviour active systems \cite{araujo2023steering}. Therefore we investigate the interaction between activity and viscoelasticity by analyzing an active nematic flowing in a soft channel.

Previous studies that address the role of viscoelasticity in living systems considered either active particles within a viscoelastic fluid \cite{juelicher2007active,marcq2014spatio,li2016collective,hemingway2016viscoelastic,liu2021viscoelastic,de2023pulsatory} or active matter in contact with a viscoelastic environment \cite{emmanuel2020active,plan2021activity,emmanuel2022self}. In the former case, oscillating vortices and drag reduction effects are seen to arise due to the presence of polymers \cite{liu2021viscoelastic,hemingway2016viscoelastic}. In the latter, less studied case, numerical simulations demonstrate that temporal pulses in activity drive reversal of spontaneous flows \cite{plan2021activity}. In this \textit{letter}, we demonstrate analytically and numerically that, above a critical activity, viscoelastic confinement produces spontaneous, oscillatory flow states of an active nematic that switches flow directions periodically. The direction-reversing oscillatory flows exist only in `soft' channels, and they disappear when the elastic modulus of the confinement increases above a critical value. Building on our findings, we explain the origin of oscillations as the interplay of activity and viscoelasticity, demonstrate the generality of the phenomenon and discuss the consequences.

\begin{figure}
    \centering
    \includegraphics[scale=1]{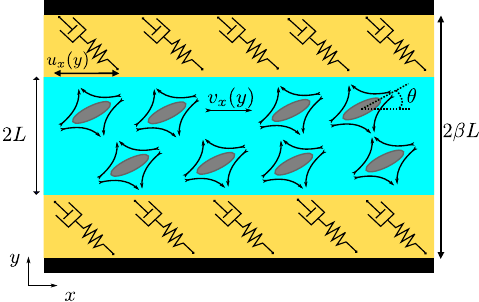}
    \caption{Schematic representation of the system: an active nematic layer of width $2L$ is confined between two viscoelastic layers, each of width $(\beta-1)L$. Thus the bounding rigid plates are separated by a distance $2\beta L$. The active nematic is a dense suspension of elements that generate active stress. The viscoelastic layers are shown as made up of Maxwell elements.}
    \label{fig:scheme}
\end{figure}

We consider a two-dimensional channel of width $2L$ and infinite length which contains the active nematic. The borders of the channel which span a width of $(\beta-1)L$ on either side are made up of viscoelastic material (see Fig.~\ref{fig:scheme}). Let $x$ and $y$ denote the directions parallel and perpendicular to the channel length, with $y=0$ the centerline of the channel. The relevant hydrodynamic variables are $\textbf{Q}$ and $\bm{v}$ representing the orientational order and velocity field in the active nematic respectively and $\bm{u}$ the displacement field in the viscoelastic layers.

Active nematics may develop orientational order either due to the elongated shape of the constituents, \cite{sanchez2012spontaneous,galanis2010nematic} or as an emergent feature of deformability of particles, such as cells \cite{mueller2019emergence} or due to activity itself \cite{santhosh2020activity}. The nematic order is measured using an orientational order parameter $\textbf{Q}=2q(\textbf{n}\textbf{n}-\textbf{I}/2)$, where $\textbf{n}=(\cos(\theta),\sin(\theta))$ is the director field, $\theta\in (-\pi/2,\pi/2)$ is the angle that the nematogens form with the positive-$x$ direction, $q$ is the magnitude of the nematic order and $\mathbf{I}$ is the identity tensor. The nematic order parameter tensor evolves according to \cite{beris1994thermodynamics}
\begin{equation}
    \left(\partial_t+\bm{v}\cdot \nabla \right)\textbf{Q}=\textbf{S}+\gamma^{-1}\textbf{H}\,,
    \label{eq:beris}
\end{equation}
where $\textbf{S}=2\lambda q\bm{\mathcal{E}}+\bm{\Omega}\cdot\bm{Q}-\bm{Q}\cdot\bm{\Omega}$ describes the generalised corotational derivative, $\bm{\mathcal{E}}=((\nabla \bm{v})^\intercal+(\nabla \bm{v}))/2$ is the strain rate tensor, and $\bm{\Omega}=((\nabla \bm{v})^\intercal-(\nabla \bm{v}))/2$ is the vorticity tensor. The flow aligning parameter $\lambda$ is determined by the shape of the nematogens. In Eq.~\eqref{eq:beris}, $\gamma$ is the rotational viscosity and $\bm{H}=-\delta\mathcal{F}/\delta \bm{Q}$ is the molecular field which drives the system to the minimum of the free energy with energy density $\mathcal{F}=\frac12 A\bm{Q}^2+\frac14 C \bm{Q}^4+\frac12 K (\nabla \bm{Q})^2$. Here, $K$ is the elastic constant, and $A$ and $C$ are material parameters, chosen so that the system is in the nematic phase at equilibrium. 

The velocity field $\bm{v}$ obeys the incompressible Navier-Stokes equations \cite{marchetti2013hydrodynamics,doostmohammadi2018active}:
\begin{equation}
\nabla\cdot\bm{v}=0\,,\quad\quad\rho_1(\partial_t \bm{v}+\bm{v}\cdot\nabla\bm{v})=\nabla\cdot\bm{\sigma}\,,\label{eq:NSE}
\end{equation}
where the total stress tensor $\bm{\sigma}$ is given by the sum of (i) the viscous stress $\bm{\sigma}^{\rm viscous}=2\eta_1\bm{\mathcal{E}}$, where $\eta_1$ is the viscosity of the active nematic, (ii) the elastic stress $\bm{\sigma}^{\rm elastic}=-P_1\bm{I}-2\lambda q \bm{H}+\bm{Q}\cdot\bm{H}-\bm{H}\cdot\bm{Q}$, where $P_1$ is the bulk pressure, and (iii) the active stress $\bm{\sigma}^{\rm active}=-\zeta \bm{Q}$. Here $\zeta$ is the activity coefficient, with $\zeta>0$ ($\zeta<0$) corresponding to extensile (contractile) activity.

The dynamics of the incompressible viscoelastic layers is described by the displacement field $\bm{u}$ from the equilibrium position, that evolves according to \cite{landau1986theory,joseph2013fluid} 
\begin{equation}
   \nabla\cdot\bm{u}=0\,,\quad\quad \rho_2\frac{\partial^2\bm{u}}{\partial t^2}=-\nabla P_2+\nabla\cdot\bm{\tau}\,,
    \label{eq:momentum_u}
\end{equation}
where $\rho_2$ is the gel density and $P_2$ is the bulk pressure in the viscoelastic layers. The stress tensor $\bm{\tau}$  is model dependent and we consider two simple yet powerful constitutive relations, namely the
\begin{align}
&\textnormal{(i) Maxwell model:}  \quad  \frac{1}{E}\frac{D\bm{\tau}}{D t}+\frac{1}{\eta_2}\bm{\tau}=\nabla\partial_t \bm{u}+(\nabla \partial_t\bm{u})^\intercal\,,\nonumber\\
& \textnormal{(ii) Kelvin-Voigt model:} \quad \bm{\tau}=(E+\eta_2 \partial_t)\nabla \bm{u}\,,\nonumber
\end{align}
to capture the rheological response of the viscoelastic layers that confine the active nematic. In the above,
$D/Dt$ is the upper convected derivative \cite{joseph2013fluid}, $E$ and $\eta_2$ are the elastic modulus and viscosity respectively. A Maxwell (Kelvin-Voigt) material is composed of a spring and a dashpot connected in series (parallel). It behaves as an elastic solid at short (long) times and as a viscous liquid at long (short) times, with a single crossover timescale $\eta_2/E$.

Eqs.~\eqref{eq:beris}-\eqref{eq:momentum_u} govern the dynamics of the system and we solve them (i) analytically as a linear stability problem and (ii) numerically using a hybrid lattice Boltzmann method \cite{supmat}. We assume translational invariance in the $x$-direction, so that $v_y=0$; $v_x=v_x(y)$ and $u_y=0$; $u_x=u_x(y)$. The viscoelastic material is in contact with a no-slip wall at $y=\pm \beta L$. At the interface between the active nematics and the viscoelastic layer, we impose no-slip conditions, $v_x(\pm L)=\partial_t u_x(\pm L)$, and continuity of the stress tensor $\sigma_{xy}(\pm L)=\tau_{xy}(\pm L)$. For simplicity, we consider strong planar anchoring of the director field at the interface, i.e., $\theta(\pm L)=0$. 

\begin{figure}
    \centering
    \includegraphics[scale=0.75]{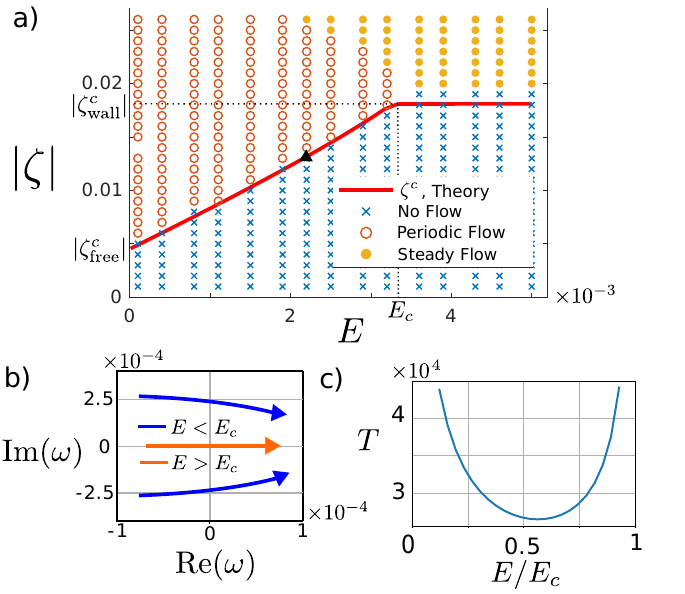}
    \caption{{\textbf a)}: Phase diagram in the $(\zeta,E)$ plane, illustrating the states of an active nematic when confined in a soft channel. The continuous red line is the critical activity $\zeta^c$, obtained from linear stability analysis, at which the nematic state becomes unstable, driving flows. The symbols are obtained from hybrid lattice Boltzmann simulations with $\eta_1=10/3$, $\eta_2=\infty$ (elastic limit), $\gamma=10$, $\rho_1=20$, $\rho_2=0$, $K=0.1$, $q_0=0.25$, $L=10$, $\lambda=0$, $\beta=2$.  For $E<E_c\approx 0.00326$, the instability leads to periodic oscillations. For $E>E_c$, a steady flow of active nematic is obtained. {\textbf b)}: The growth rate $\omega$ in the complex plane, for $E=0.002<E_c$ (blue lines) and $E=0.0042>E_c$ (orange line). The arrows indicate the direction of increasing $|\zeta|$. {\textbf c)}: Time period of oscillations $T$ as a function of $E/E_c$. $T$ diverges when $E\to 0$ and $E\to E_c$. }
    \label{fig:phase}
\end{figure}

To investigate the interplay of activity and viscoelasticity, we perform linear analysis to calculate the stability of a small perturbation around the static nematic state with $(v_x,u_x,\theta,q)=(0,0,0,q_0)$ where $q_0=\sqrt{-A/(2C)}$. For each field, we consider small perturbations around the steady state $f_0$ of the type $f(y,t)=f_0+\tilde{f}(y)e^{\omega t}$. In the limit of zero inertia ($\rho_1=\rho_2=0)$, the growth rate $\omega$ satisfies the transcendental equation \cite{supmat}
\begin{align}
         \frac{\omega+(\gamma^{-1}K\Lambda_1-\omega/\Lambda_1)\tanh(\Lambda_1L)}{(\eta_1\omega+q_0(1-\lambda)\zeta)L}
    =\frac{(1-\beta)}{E \mathcal{T}}\,
      \label{eq:condition_omega}
\end{align}
where $\Lambda_1=\sqrt{\frac{\eta_1\omega+q_0(1-\lambda)\zeta}{\eta_1\gamma^{-1}K+2q_0^2K(\lambda-1)^2}}$. For Maxwell and Kelvin-Voigt models, respectively, $\mathcal{T} = (\omega + E/\eta_2)^{-1}$ and $\mathcal{T} = \omega^{-1}+\eta_2/E$. The nature of instability associated with the system depends on $\omega$, the solution of Eq.~\eqref{eq:condition_omega}. 

For simplicity, we first consider a purely elastic material bounding the nematic fluid, corresponding to the limit $\eta_2\to\infty$ ($\eta_2\to 0$) for the Maxwell (Kelvin-Voigt) model. In the limit of large elastic modulus $E\to\infty$, the boundaries at $y=\pm L$ are rigid and we recover the classical result of \citet{voituriez2005spontaneous}: increasing the activity beyond a critical value $\zeta^c_{\rm wall}$, the nematically ordered state is unstable and spontaneous flows develop driven by the distortions in the director field. The critical activity is calculated from Eq.~\eqref{eq:condition_omega},
\begin{equation}
    \zeta^c_{\rm wall}=-\frac{\pi^2K\left[\eta_1/\gamma+2q_0^2(1-\lambda)^2\right]}{2q_0(1-\lambda)L^2}\,.
\end{equation}
In the opposite limit $E=0$, corresponding to a free-standing film of active nematic, an analogous transition to a steady flow is observed at activity $\zeta^c_{\rm free}=\zeta^c_{\rm wall}/4$. 

The critical activity of the system at intermediate values of $E$, obtained from Eq.~\eqref{eq:condition_omega}, is summarized in Fig.~\ref{fig:phase} (red line). The critical activity $\zeta^c=\zeta^c_{\rm free}$ at $E = 0$, and increases with increase in the elastic modulus $E$, until a threshold elastic modulus $E = E_c$. Beyond $E_c$ the critical activity ``freezes'' to $\zeta^c=\zeta^c_{\rm wall}$, that corresponding to a rigid wall. $E_c$ can be determined analytically from Eq.~\eqref{eq:condition_omega}, see \cite{supmat}. 

Interestingly the transition mechanism at $\zeta^c$, at which the ordered nematic state becomes unstable, is different for $E<E_c$ and $E>E_c$. We find that, for $E < E_c$ the route to instability is via a Hopf bifurcation where the complex conjugate eigenvalues $\omega$ cross the imaginary axis with a finite imaginary part at $\zeta=\zeta^c$ (Fig.~\ref{fig:phase}b). Consequently, the ensuing instability is oscillatory and the active nematic transitions from a quiescent to an oscillating state where the flow direction is reversed periodically. On the other hand, for $E > E_c$, the instability becomes stationary ($\textnormal{Im}(\omega) = 0$) and no oscillations are observed. The numerical simulations show that the oscillations are replaced by steady flow at sufficiently high activity (see Fig.~\ref{fig:phase}).

\begin{figure}
    \centering
    \includegraphics[width=0.48\textwidth]{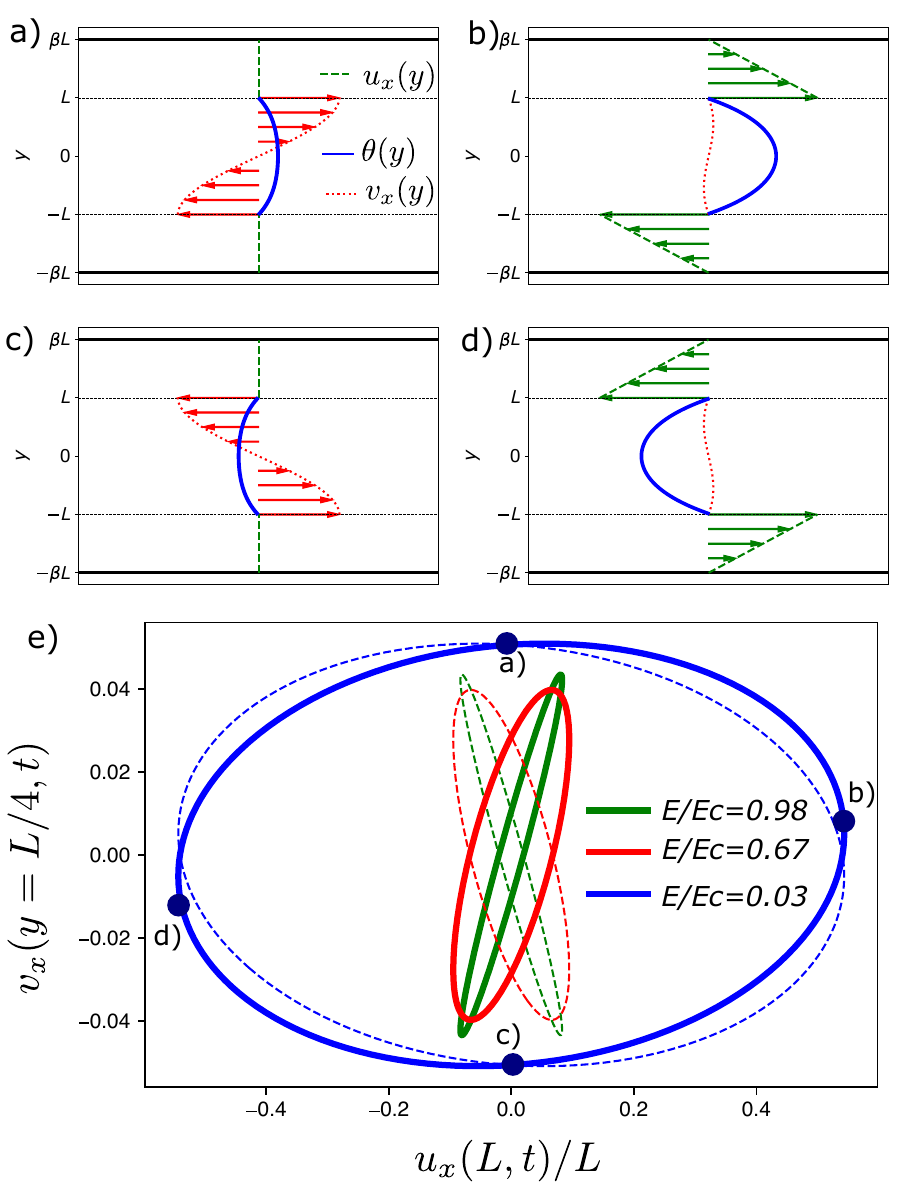}
        \caption{Temporal evolution of the states of the system with the same parameter values as in Fig.~\ref{fig:phase} at $\zeta=\zeta^c(E)$.  The panels \textbf{a)}, \textbf{b)}, \textbf{c)}, and \textbf{d)} show the hydrodynamic fields in the oscillatory phase for $t=0$, $t=T/4$, $t=T/2$, and $t=3T/4$. Panel \textbf{e)} shows the phase space trajectory in the $(u_x(L,t),v_x(L/4,t))$ plane for different values of $E$. The dashed lines display the time-reversed trajectories $(u_x(L,T-t),-v_x(L/4,T-t))$, showing time-irreversibility.}
    \label{fig:trajectories}
\end{figure}

The oscillatory state can be understood by following the temporal evolution of a system which is at its critical activity $\zeta=\zeta^c$, and with $0<E<E_c$ and $|\zeta_{\rm free}^c|<|\zeta|<|\zeta_{\rm wall}^c|$ (such as a point marked `$\blacktriangle$' in Fig.~\ref{fig:phase}(a)). At time $t=0$ (see Fig.~\ref{fig:trajectories}a), the elastic layer is  not deformed ($u_x=0$), and the stress at the active nematic-elastic interface  ($y=\pm L$) vanishes. This condition corresponds to a free standing active nematic film (no resistance from the elastic layer), which will have a critical activity $\zeta^c_{\rm free}$. Since the activity of the system exceeds this critical value, $|\zeta|>|\zeta^c_{\rm free}|$, spontaneous flow develops in the active film. The velocity profile $\tilde{v}_x(y)$ is an odd function of $y$ \cite{supmat} similar to that of a shear flow. These flows, in turn, drive the deformation of the elastic confinement. Eventually, the elastic response of the channel wall slows down the flow and the deformation rate at the active-elastic interface vanishes. In this configuration, the effect of elastic confinement is the same as that of a rigid wall and the critical activity for the active nematic is $\zeta^c_{\rm wall}$. However, since $|\zeta|<|\zeta^c_{\rm wall}|$, the active forcing is not sufficient to sustain the flows and they die out (Fig.~\ref{fig:trajectories}b). The elastic energy stored in the elastic medium pushes the flow in the opposite direction, leading to a flow reversal (Fig.~\ref{fig:trajectories}c). Hence, the oscillations arise because the activity is too high to remain in the quiescent state ($\zeta > \zeta^c_{\rm free}$) but too low to sustain the flow ($\zeta < \zeta^c_{\rm wall}$).

The period $T$ of the oscillations is set by the elasticity, viscosity and $L/(\gamma^{-1}K)$, the relaxation timescale of the director field.
The period $T$ close to the critical point $\zeta=\zeta^c$ can be obtained analytically \cite{supmat} from Eq.~\eqref{eq:condition_omega} and is shown in Fig.~\ref{fig:phase}c as a function of $E$. For $E\to 0$, the activity $\zeta$ is only slightly larger than $\zeta_{\rm free}^c$ required to initially start a flow, leading to a slowdown of the dynamics. Similarly, when $E\to E_c$, the activity $\zeta$ is only marginally below $\zeta_{\rm wall}^c$ and the flow-reversal mechanism again slows down significantly. Indeed the time period diverges in the limiting cases: $T\sim \sqrt{\frac{L\eta_1}{\gamma^{-1}KE}}$ for $E\to 0$ and $\sqrt{\frac{L\eta_1}{\gamma^{-1}K(E_c-E)}}$ for $E\to E_c$. Hence, the crossover from oscillatory to steady flow at the two limiting cases, $E > 0$ to $E=0$ and $E < E_c$ to $E=E_c$ occurs smoothly via an infinite-period bifurcation. The period $T$ has a minimum at $E = E^*$, reminiscent of the phenomenon of resonance and the elastic modulus can be optimally tuned to increase the frequency of  oscillatory motion. 

To gain further insight into the oscillatory modes of the instability, we next plot the trajectory of the system in a phase space spanned by the displacement of the elastic layer ($u_x(y=L,t)$) and the velocity of the active nematic ($v_x(y=L/4,t)$) as shown in Fig.~\ref{fig:trajectories}e. The exact shape of the curve depends on the choice of parameters, but note that the phase space trajectory encloses a finite area 
indicating the phase lag in the  the velocity field of active nematic and the displacement field of elastic confinement. Interestingly, the phase space trajectory does not coincide with the time-reversed trajectory $(u_x(L,-t),-v_x(L/4,-t))$, manifestly breaking the time-reversal symmetry and showing the non-equilibrium nature of the active-dissipative system under consideration. While non-reciprocal oscillatory motion, the sine qua non for self-propulsion (the scallop theorem), is abundant in life at low Reynolds number \cite{purcell1977life,lauga2011life}, the current analysis demonstrates that the mechanical coupling of activity and elasticity automatically generates such non-reciprocal motion in active systems.

\begin{figure}
    \centering
    \includegraphics[scale=0.4]{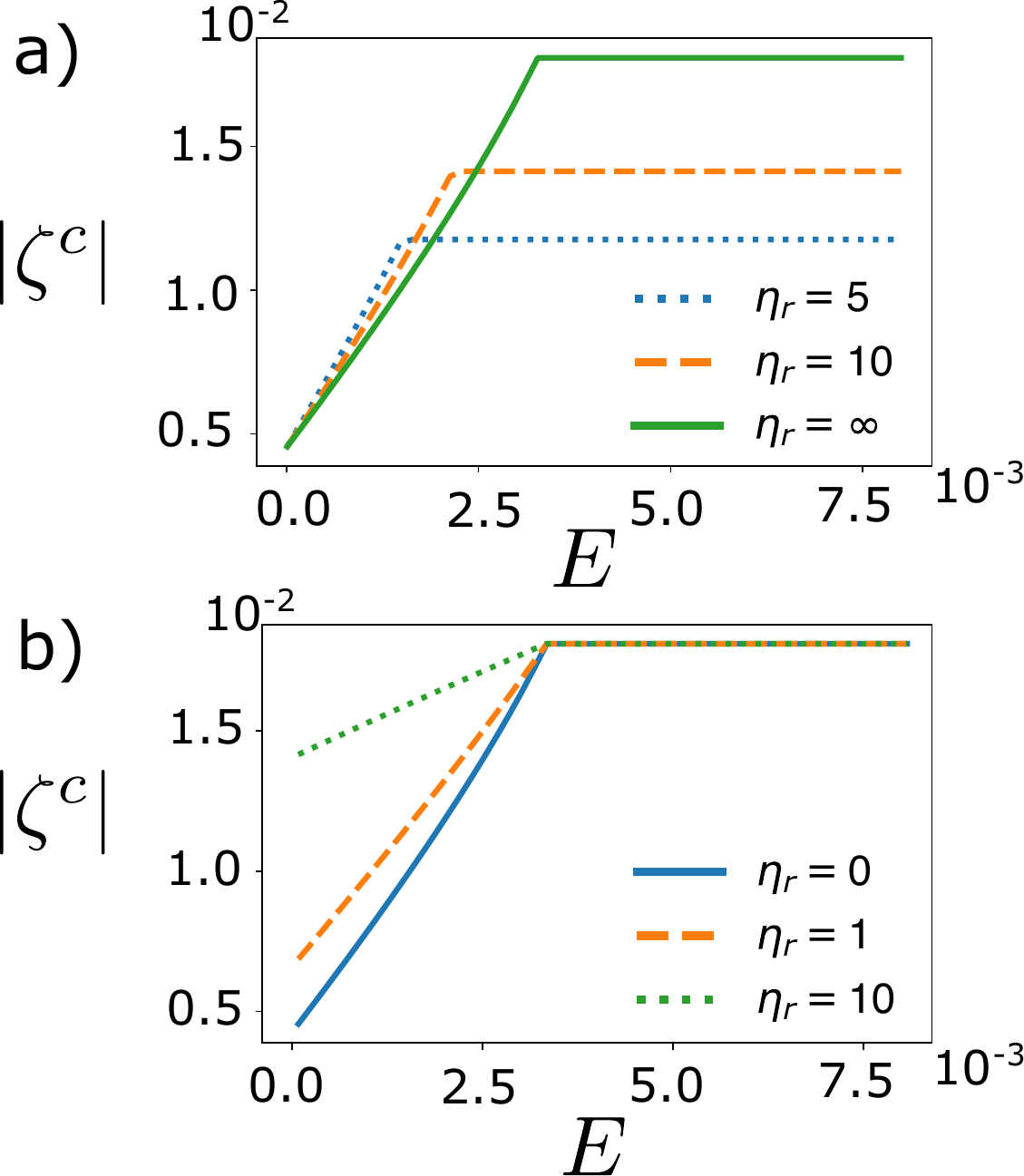}
    \caption{Critical activity $|\zeta^c|$ as a function of the elastic modulus $E$ for different values of $\eta_r=\eta_2/\eta_1$ for the Maxwell model (panel \textbf{a}) and the Kelvin Voigt model (panel \textbf{b}). For both panels, the values of the parameters are $\eta_1=10/3$, $\gamma=10$, $\rho_1=\rho_2=20$, $K=0.1$, $q_0=0.25$, $L=10$, $\lambda=0$, $\beta=2$. In the regions where $|\zeta^c|$ increases with $E$ the instability is oscillatory. 
    }
    \label{fig:crit_zeta}
\end{figure}

Having established that the genesis of oscillations is the elasticity of the confining channel we can analyze more complex constitutive relations. For the Maxwell model, on time scales smaller than $\eta_2/E$ the viscoelastic confinement behaves as an elastic solid and the coupling between activity and elasticity still leads to oscillations as illustrated in Fig.~\ref{fig:crit_zeta}(a). The instability becomes stationary for $E>E_c$ but with the difference that $E_c$ depends on the viscosity ratio $\eta_r = \eta_2/\eta_1$. When $E>E_c$, the viscoelastic confining material essentially behaves as a viscous fluid with critical activity $\zeta^c_{\rm visc}$. Hence, the critical activity $\zeta^c=\zeta^c_{\rm visc}$ depends upon the viscosity ratio $\eta_r$ and $|\zeta^c_{\rm free}|<|\zeta^c_{\rm visc}|<|\zeta^c_{\rm wall}|$. The behaviour at small $E$ can  be understood in a similar fashion. In this limit, the viscoelastic timescale $\eta_2/E$ is large compared to the period of the oscillations $T\sim 1/\sqrt{E}$, and the Maxwell material behaves as an elastic solid exhibiting an $\eta_r$ independent behaviour of $\zeta^c$. In particular, $\zeta^c\approx\zeta^c_{\rm free}$ at small elasticity $E$.

Opposite behaviours are observed when the channel confinement is the Kelvin-Voigt material. For $E>E_c$ the instability is still stationary but the confining material now behaves as an elastic solid and $\zeta^c=\zeta^c_{\rm wall}$. This results in the threshold elastic modulus $E_c$ being independent of the viscosity $\eta_2$ as illustrated in Fig.~\ref{fig:crit_zeta}(b). On the other hand, for $E\to 0$, the viscoelastic timescale $\eta_2/E$ is large compared to the period of the oscillations and the Kelvin-Voigt material behaves as a viscous fluid. Hence, the critical activity $\zeta^c$ for small $E$ strongly depends on the viscosity ratio $\eta_r$. To summarize, choice of different constitutive models of the channel confinement leads to quantitative differences but does not change the physics of the oscillations.

Our results highlight a novel pathway to spatiotemporal pattern formation in active matter. It is indeed remarkable to note that the time periodic, oscillatory flows arise even at constant activity. Our predictions can be tested experimentally, by confining cell layers \cite{duclos2018spontaneous} or microtubule-based active fluids \cite{chandrakar2020confinement} in channels with soft walls. Moreover, traction force microscopy provides a potential platform to study the role of an active-elastic boundary \cite{style2014traction,colin2018future}. In addition to extracting work from active materials, the nonreciprocal dynamics that arise from the interplay of activity and viscoelasticity might also be utilised to make self propelling objects. 

This work was supported by a Leverhulme Trust International Professorship Grant (No.~LIP-2020-014). SB acknowledges support from the Rhodes Trust and the Crewe Graduate Award.

\bibliographystyle{apsrev4-1} 
\bibliography{biblio} 

\newpage
\onecolumngrid

\section{Supplemental Material}

\subsection{Linear stability analysis}

In this section, we perform the linear stability analysis of the Eqs.~(\ref{eq:beris}-\ref{eq:momentum_u}). Under the assumption of translational invariance in the $x$ direction, the governing equations for the nematic region ($|y|<L$) become
\begin{equation}
    \begin{cases}
        \partial_t q&=\gamma^{-1}\left[-q\left(A+2Cq^2-\gamma\lambda\sin(2\theta)\partial_y v_x+4K(\partial_y\theta)^2\right)+K\partial^2_y\theta\right]\,,\\
        \partial_t \theta&=2\gamma^{-1}K q^{-1} \partial_y q\partial_y\theta+\frac12 \gamma^{-1}\left[\gamma\partial_y v_v(\lambda\cos(2\theta)-1)+2K\partial^2_y\theta\right]\\
        \rho_1\partial_tv_x&=\partial_y\sigma_{xy}\,,
    \end{cases}
\end{equation}
where
\begin{align}
    \sigma_{xy}&=-2\lambda q\left[\sin(2\theta)q(-A-2B\cos^2(2\theta)q^2)+K(4\partial_yq\partial_y\theta \cos(2\theta)-4q\sin(2\theta)(\partial_y\theta)^2)\right]+4Kq\left[2\partial_yq\partial_y\theta+q\partial^2_y\theta\right]\nonumber \\
    &+\eta_1\partial_yv_x-\zeta q\sin(2\theta)\,.
\end{align}
In the viscoelastic region ($y>|L|)$, we find
\begin{equation}
    \begin{cases}
     & \rho_2\partial^2_tu_x=\partial_y\tau_{xy}\,,\\
       & \frac{1}{E}\frac{D\tau_{xy}}{D t}+\frac{1}{\eta_2}\tau_{xy}=\partial_y\partial_tu_x \,,
    \end{cases}
\end{equation}
where $D\tau_{xy}/(Dt)=\partial_t\tau_{xy}-\tau_{yy}\partial_y\partial_t u_x$ is the upper-convected derivative of the stress tensor. The boundary conditions are 
\begin{equation}
    \theta(y=\pm L)=0\,,\quad\quad v_x(y=\pm L )=\partial_t u_x(y=\pm L)\,,\quad\quad\tau_{xy}(\pm L)=\sigma_{xy}(\pm L)\,,\quad\quad u_x(\pm\beta L)=0\,.
\end{equation}

We probe the stability of a small perturbation around the stationary state $(v_x,u_x,\theta,q)=(0,0,0,q_0=\sqrt{-A/(2C)})$ of the type $f(y,t)=f_0+\tilde{f}(y)e^{\omega t}$. Expanding to linear order, we find
\begin{equation}
    \begin{cases}
       \omega \tilde{\theta}&=\gamma^{-1}K\partial^2_{y}\tilde{\theta}+\frac{\lambda-1}{2}\partial_y\tilde{v}_x\,,\\
        \rho_1\omega\tilde{v}_x&=\partial_y\sigma_{xy}\,,
        \label{eq:linear_v_theta}
    \end{cases}
\end{equation}
where
\begin{equation}
    \sigma_{xy}=\eta_1\partial_y\tilde{v}_x-2\zeta q_0\tilde{\theta}-4q_0^2 K(\lambda-1)\partial_y^2\tilde{\theta}\,,\label{eq:linear_theta_v}
\end{equation}
and
\begin{equation}
\partial^2_y\tilde{u}_x= \delta^2\tilde{u}_x\,,\label{eq:gel_eq}
\end{equation}
where
\begin{equation}
   \delta= \sqrt{\rho_2\omega\left(\frac{\omega}{E}+\frac{1}{\eta_2}\right) }\,.
\end{equation}
The boundary conditions are
\begin{align}
&\eta_1\partial_y\tilde{v}_x(\pm L)-2\zeta q_0\tilde{\theta}(\pm L)-4q_0^2K(\lambda-1)\partial_y^2\tilde{\theta}(\pm L)=\frac{\omega}{\omega/E+1/\eta_2} \partial_y\tilde{u}_x(\pm L)\,,\\
       &\tilde{\theta}(\pm L)=0\,,\quad  \tilde{v}_x(\pm L )=\omega  \tilde{u}_x(\pm L)\,,\quad \tilde{u}_x(\pm\beta L)=0\,.
\end{align}

Solving the gel equation \eqref{eq:gel_eq} and imposing the no-slip boundary condition $\tilde{u}_x(\pm \beta L)=0$ we find
\begin{equation}
    u_x(y)=c_4^{\pm}\left[\sinh(\delta y)\mp \tanh(\delta\beta L)\cosh(\delta y)\right]\,,
    \label{eq:ux}
\end{equation}
for $y>L$ ($y<-L)$. From Eq.~\eqref{eq:linear_theta_v}, we obtain 
\begin{equation}
\partial^4_y\tilde{\theta}(y)-a\partial^2_y\tilde{\theta}(y)+b\tilde{\theta}(y)=0\,,
\label{eq:diff_lin_theta}
\end{equation}
where we have defined 
\begin{equation}
a=\frac{\eta_1\omega+\gamma^{-1}K\rho_1\omega-q_0\zeta (\lambda-1)}{\eta_1\gamma^{-1}K+2q_0^2K(\lambda-1)^2}\,,
\end{equation}
and
\begin{equation}
    b=\frac{\rho_1 \omega^2}{{\eta_1\gamma^{-1}K+2q_0^2K(\lambda-1)^2}}\,.
\end{equation}
We first consider the even solution
\begin{equation}
   \theta(y)=c_1\left[\cosh\left(\Lambda_1 y\right)-\frac{\cosh(\Lambda_1L)}{\cosh(\Lambda_2L)}\cosh\left(\Lambda_2 y\right)\right]\,,
\end{equation}
where we have imposed the boundary condition $\theta(\pm L)=0$ and defined
\begin{equation}
    \Lambda_{1,2}=\sqrt{\frac{a\pm \sqrt{a^2-4b}}{2}}\,.
\end{equation}
Using Eq.~\eqref{eq:linear_v_theta}, we find
\begin{equation}
    \tilde{v}_x(y)=\frac{2}{1-\lambda}c_1\left[\left(\gamma^{-1}K\Lambda_1-\frac{\omega}{\Lambda_1}\right)\sinh(\Lambda_1 y)-\left(\gamma^{-1}K\Lambda_2-\frac{\omega}{\Lambda_2}\right)\frac{\cosh(\Lambda_1 L)}{\cosh(\Lambda_2 L)}\sinh(\Lambda_2 y)\right]\,.
\end{equation}
Imposing the boundary conditions at $y=\pm L$, we find the following condition for $\omega$
\begin{align}
     \frac{\left(\gamma^{-1}K\Lambda_1-\omega/\Lambda_1\right)\tanh\left(\Lambda_1 L\right)-\left(\gamma^{-1}K\Lambda_2-\omega/\Lambda_2\right)\tanh\left(\Lambda_2L \right)}{(\Lambda_1^2-\Lambda_2^2)\left[\gamma^{-1}\eta_1 K+2q_0^2 K(1-\lambda)^2\right]}
    +\left(\frac{\omega}{E}+\frac{1}{\eta_2}\right)\frac{\tanh\left(\delta(\beta-1)L\right)}{\delta}=0\,.
    \label{eq:condition_nonet}
\end{align}
Considering the odd solution, we find
\begin{equation}
   \theta(y)=c_1\left[\sinh\left(\Lambda_1 y\right)-\frac{\sinh(\Lambda_1L)}{\sinh(\Lambda_2L)}\sinh\left(\Lambda_2 y\right)\right]\,,
\end{equation}
\begin{equation}
    \tilde{v}_x(y)=\frac{2}{1-\lambda}c_1\left[\left(\gamma^{-1}K\Lambda_1-\frac{\omega}{\Lambda_1}\right)\cosh(\Lambda_1 y)-\left(\gamma^{-1}K\Lambda_2-\frac{\omega}{\Lambda_2}\right)\frac{\sinh(\Lambda_1 L)}{\sinh(\Lambda_2 L)}\cosh(\Lambda_2 y)\right]\,,
\end{equation}
and the condition
\begin{align}
     \frac{\left(\gamma^{-1}K\Lambda_1-\omega/\Lambda_1\right)\coth\left(\Lambda_1L \right)-\left(\gamma^{-1}K\Lambda_2-\omega/\Lambda_2\right)\coth\left(\Lambda_2L \right)}{(\Lambda_1^2-\Lambda_2^2)\left[\gamma^{-1}\eta_1 K+2q_0^2 K(1-\lambda)^2\right]}
    +\left(\frac{\omega}{E}+\frac{1}{\eta_2}\right)\frac{\tanh\left(\delta(\beta-1)L\right)}{\delta}=0\,.
    \label{eq:condition_net}
\end{align}
For the range of parameters considered in the paper, we find that the even solution (corresponding to no net flow in the channel) is dominant, i.e., it becomes unstable at lower values of the activity. Hence, in the main text, we only focus on the even mode. The odd solutions may be favored by introducing weak anchoring.

\subsection{Asymptotic behaviors}

In this section, we extract the asymptotic behavior of the solution of Eq.~\eqref{eq:condition_nonet}. For simplicity, we set $q_0=1$ and we consider the case $\rho_1=\rho_2=0$. In this limit, the condition in Eq.~\eqref{eq:condition_nonet} becomes
\begin{align}
     \frac{\omega+(\gamma^{-1}K\Lambda_1-\omega/\Lambda_1)\tanh(\Lambda_1L)}{\eta_1\omega+(1-\lambda)\zeta}
    +\left(\frac{\omega}{E}+\frac{1}{\eta_2}\right)(\beta-1)L=0\,,
    \label{eq:condition_nonet_limit}
\end{align}
where
\begin{equation}
    \Lambda_1=\sqrt{\frac{\eta_1\omega+ (1-\lambda)\zeta}{\eta_1\gamma^{-1}K+2K(\lambda-1)^2}}\,.
\end{equation}
We first consider the limit of small $E$. For $E=0$ (corresponding to a free surface), the critical value of the activity can be computed analytically and reads
\begin{equation}
    \zeta^c_{\rm free}=-\pi^2\frac{\eta_1\gamma^{-1}K+2K(1-\lambda)^2}{4L^2 (1-\lambda)}.
\end{equation}
We set $E=\epsilon$, $\zeta=\zeta^c_{\rm free}+a_1\epsilon$, and $\omega=a_2\sqrt{\epsilon}$. We then expand Eq.~\eqref{eq:condition_nonet_limit} in powers of $\epsilon$, yielding
\begin{align}
\frac{{a_2 L (-1 + \beta) + \frac{{2 K}}{{a_2 \gamma \eta_1}}}}{{\sqrt{\epsilon}}} + \left[\frac{{L (-1 + \beta)}}{{\eta_2}} + \frac{{2 \left(\frac{{\eta_1 (5 \eta_1 + 8   \gamma (-1 + \lambda)^2)}}{{\pi^2 (\eta_1 + 2   \gamma (-1 + \lambda)^2)}} + \frac{{a_1 K   (-1 + \lambda)}}{{a_2^2 \gamma}}\right)}}{{\eta_1^2}} \right]+\mathcal{O}(\sqrt{\epsilon})=0\,.
\end{align}
Setting the coefficients to zero, we get
\begin{equation}
    a_2=\pm i\frac{\sqrt{2\gamma^{-1}K}}{\sqrt{L\eta_1(\beta-1)}}
\end{equation}
and
\begin{equation}
   a_1= -\frac{{ \eta_1^2 \left(\frac{{L (-1 + \beta)}}{{\eta_2}} + \frac{{2 (5 \eta_1 + 8   \gamma (-1 + \lambda)^2)}}{{\pi^2 \eta_1 (\eta_1 + 2   \gamma (-1 + \lambda)^2)}}\right)}}{{  L  \eta_1 ( \beta-1)  (1- \lambda)}}\,.
\end{equation}
As expected, the growth rate $\omega$ is purely imaginary.

To investigate the asymptotic behavior of the system close to the transition, we set $E=E_c-\epsilon$, $\zeta=\zeta^c_{\rm visc}-a_1\epsilon$, and $\omega=a_2\sqrt{\epsilon}$, yielding
\begin{align}
&\left[\frac{{L ( \beta-1)}}{{\eta_2}} - \frac{{K \sqrt{-\frac{{  \gamma \zeta^c_{\rm visc} (-1 + \lambda)}}{{K (\eta_1 + 2   \gamma (-1 + \lambda)^2)}}}} \tanh\left(\sqrt{-\frac{{  \gamma \zeta^c_{\rm visc} (-1 + \lambda)}}{{K (\eta_1 + 2   \gamma (-1 + \lambda)^2)}}}\right)}{{-  \gamma \zeta^c_{\rm visc} +   \gamma \zeta^c_{\rm visc} \lambda}}\right]\\
&+\left[\frac{{(a_2 L (-1 + \beta))}}{{Ec}} - \frac{{a_2 (3 \eta_1 + 4   \gamma (-1 + \lambda)^2)}}{{2   \zeta^c_{\rm visc} (\eta_1 + 2   \gamma (-1 + \lambda)^2) (-1 + \lambda)}}+ \frac{{a_2 \eta_1 \tanh\left(\sqrt{-\frac{{  \gamma \zeta^c_{\rm visc} (-1 + \lambda)}}{{K (\eta_1 + 2   \gamma (-1 + \lambda)^2)}}}\right)^2}}{{2   \zeta^c_{\rm visc} (\eta_1 + 2   \gamma (-1 + \lambda)^2) (-1 + \lambda)}}\right. \\  &- \left. \frac{{a_2 K (3 \eta_1 + 4   \gamma (-1 + \lambda)^2) \sqrt{-\frac{{  \gamma \zeta^c_{\rm visc} (-1 + \lambda)}}{{K (\eta_1 + 2   \gamma (-1 + \lambda)^2)}}} \tanh\left(\sqrt{-\frac{{  \gamma \zeta^c_{\rm visc} (-1 + \lambda)}}{{K (\eta_1 + 2   \gamma (-1 + \lambda)^2)}}}\right)}}{{2 q \gamma (\zeta^c_{\rm visc})^2 (-1 + \lambda)^2}} \right] \sqrt{\epsilon}+\mathcal{O}(\epsilon)=0\,.
\end{align}
Setting the coefficients to zero we find
\begin{equation}
    \frac{{L ( \beta-1)}}{{\eta_2}} - \frac{{K \sqrt{-\frac{{ \gamma \zeta^c_{\rm visc} (-1 + \lambda)}}{{K (\eta_1 + 2  \gamma (-1 + \lambda)^2)}}}} \tanh\left(\sqrt{-\frac{{ \gamma \zeta^c_{\rm visc} (-1 + \lambda)}}{{K (\eta_1 + 2  \gamma (-1 + \lambda)^2)}}}\right)}{{- \gamma \zeta^c_{\rm visc} +  \gamma \zeta^c_{\rm visc} \lambda}}=0\,,
    \label{eq:trans_zc_visc}
\end{equation}
and
\begin{align}
E_c&=\left[{2 K L  (-1 + \beta) \zeta^c \eta_2^2 (\eta_1 + 2  \gamma (-1 + \lambda)^2) (-1 + \lambda)}\right]/\left[K \eta_2^2 (3 \eta_1 + 4  \gamma (-1 + \lambda)^2)\right.\\
&+\left. K L (-1 + \beta) \eta_2 (\eta_1 + 2  \gamma (-1 + \lambda)^2) (3 \eta_1 + 4  \gamma (-1 + \lambda)^2) + L^2  (-1 + \beta)^2 \gamma \zeta^c \eta_1 (\eta_1 + 2  \gamma (-1 + \lambda)^2) (-1 + \lambda)\right]\nonumber\,.
\end{align}
Eq.~\eqref{eq:trans_zc_visc} is transcendental and must be solved numerically to determine $\zeta^c_{\rm visc}$. Considering higher order expansions, one can find expressions for $a_1$ and $a_2$. In the limit of a purely elastic medium ($\eta_2\to \infty)$ we find
\begin{equation}
   E_c=\frac{{2 K L \pi^2 (-1 + \beta) (\eta_1 + 2  \gamma (-1 + \lambda)^2)^2}}{{\gamma (3 \eta_1 + 4  \gamma (-1 + \lambda)^2)}}\,.
\end{equation}

\subsection{Kelvin-Voigt model}

In this section, we perform the linear stability analysis in the case of the Kelvin-Voigt model. The constitutive relation reads
\begin{equation}
    \tau_{xy}=(E+\eta_2\partial_t)\partial_yu_x\,.
\end{equation}
Hence, the gel displacement $u_x(y,t)$ evolves according to
\begin{equation}
    \rho_2\partial^2_t u_x=(E+\eta_2\partial_t)\partial^2_y u_x\,.
\end{equation}
Assuming $u_x(y,t)=e^{\omega t}\tilde{u}_x(y)$, we find
\begin{equation}
\partial^2_y\tilde{u}_x= \delta_{KV}^2\tilde{u}_x\,,
\end{equation}
where we have defined
\begin{equation}
    \delta_{KV}=\sqrt{\frac{\rho_2 \omega^2}{E+\eta_2\omega}}\,.
\end{equation}
Following the same derivation as for the Maxwell model, we find two instabilities, corresponding to the even and odd solutions for $\theta$. The condition for the growth rate of the even solution reads
\begin{align}
     \frac{\left(\gamma^{-1}K\Lambda_1-\omega/\Lambda_1\right)\tanh\left(\Lambda_1 \right)-\left(\gamma^{-1}K\Lambda_2-\omega/\Lambda_2\right)\tanh\left(\Lambda_2 \right)}{(\Lambda_1^2-\Lambda_2^2)\left[\gamma^{-1}\eta_1 K+2q_0 K(1-\lambda)^2\right]}+\frac{\omega}{E+\omega \eta_2}\frac{\tanh\left(\delta_{KV}(\beta-1)\right)}{\delta_{KV}}=0\,.
    \label{eq:condition_nonet_KV}
\end{align}
For the odd solution, we find
\begin{align}
     \frac{\left(\gamma^{-1}K\Lambda_1-\omega/\Lambda_1\right)\coth\left(\Lambda_1 \right)-\left(\gamma^{-1}K\Lambda_2-\omega/\Lambda_2\right)\coth\left(\Lambda_2 \right)}{(\Lambda_1^2-\Lambda_2^2)\left[\gamma^{-1}\eta_1 K+2q_0 K(1-\lambda)^2\right]}
    +\frac{\omega}{E+\omega \eta_2}\frac{\tanh\left(\delta_{KV}(\beta-1)\right)}{\delta_{KV}}=0\,.
    \label{eq:condition_net_KV}
\end{align}
As for the Maxwell model, for the range of parameters considered in the paper, we find that the even solution is dominant.

\end{document}